\documentclass[twocolumn,apl,amsmath,amssymb,showpacs,superscriptaddress]{revtex4-1}
\usepackage{epsf}      
\usepackage{graphicx}
\usepackage{color}
\usepackage{gensymb}
\usepackage{amsmath}
\usepackage{amssymb}
\usepackage{fontenc}

\begin{document}

\title {Investigation of lattice dynamics, magnetism and electronic transport in $\beta$-Na$_{0.33}$V$_2$O$_{5}$}
\author{Rishabh Shukla}
\affiliation{Department of Physics, Indian Institute of Technology Delhi, Hauz Khas, New Delhi-110016, India}
\author{Clemens Ulrich}
\affiliation{School of Physics, The University of New South Wales, Kensington, 2052 NSW, Sydney, Australia}
\author{R. S. Dhaka}
\email{rsdhaka@physics.iitd.ac.in}
\affiliation{Department of Physics, Indian Institute of Technology Delhi, Hauz Khas, New Delhi-110016, India}

\date{\today}                                     

\begin{abstract}
We investigate the electronic and magnetic properties as well as lattice dynamics and spin-phonon coupling of $\beta$-Na$_{0.33}$V$_2$O$_5$ using temperature-dependent Raman scattering, dc-magnetization and dc-resistivity, x-ray photoemission, and absorption spectroscopy. The Rietveld refinement of XRD pattern with space group C2/m confirms the monoclinic structure. The analysis of temperature-dependent Raman spectra in a temperature range of 13--673~K reveals an anharmonic dependence of the phonon frequency and full width at half maximum, which is accredited to the symmetric phonon decay. However, below about 40 K, the hardening of the phonon frequency beyond anharmonicity is attributed to the spin-phonon coupling. Interestingly, the estimated effective magnetic moment $\mu_{\rm eff}=$ 0.63~$\mu_B$ from the magnetization data manifests a mixed-valence state of V ions in 4+ (18$\pm$1\%) and 5+ (82$\pm$1\%). A similar ratio of V$^{4+}$ to V$^{5+}$ is also observed in the x-ray photoemission and x-ray absorption near-edge spectra and that is found to be consistent with the sample stoichiometry. The analysis of extended x-ray absorption fine structure at the V K-edge gives the corresponding V--O bond lengths, which are utilized in the detailed analysis of Raman modes. Moreover, the temperature-dependent resistivity resembles a semiconducting behavior where the charge carrier transport is facilitated by the band conduction at higher temperatures and via hopping $\le$260~K. 
\end{abstract}

\maketitle

\section{\noindent ~Introduction}

The sodium-based vanadium bronzes have attracted significant attention owing to their peculiar properties like complex magnetic ordering \cite{HeinrichPRL04, OhwadaPRB12}, pressure-induced superconductivity \cite{YamauchiPRL02}, nonlinear transport \cite{SirbuEPJB06}, metal-to-insulator transition accompanied with charge ordering \cite{SuzukiPRB09, YamauchiSSS05}, pressure-induced phase transition \cite{GrzechnikJPCM16}. Additionally, the layered vanadium-based compounds are considered as potential electrode materials for Li- and Na-ion batteries \cite{HadjeanJMC11, SarohaAO19, ChandraEA20}. Interestingly, the structure of vanadium bronzes is governed by the distribution of different geometric sites of V ions in the (V$_2$O$_5$)$_n$ framework, where comprehensive studies on M$_{x}$V$_2$O$_5$ have shown that the obtained phases and their homogeneity strongly depend on the value of $x$ \cite{MarleyCC15, GalyJSSC70}. As the V-ions can exist in 3+, 4+ and/or 5+ oxidation states, among which 3+ (3$d^2$) and 4+ (3$d^1$) states are magnetic, while 5+ is non-magnetic having 3$d^0$ configuration. The presence of mixed-valence states (4+ and 5+) with 1:5 ratio in $\beta$-Na$_{0.33}$V$_2$O$_{5}$ leads to the phenomenon of charge ordering (CO) as well as interesting physical properties \cite{KhooJMC10, HeinrichPRL04, YamauchiJPSJ08, ChakravertyPRB78}. Moreover, it has been reported that $\beta$-Na$_{0.33}$V$_2$O$_{5}$ exhibits a metallic behavior at high temperatures along the $b-$axis, while semiconducting behavior was observed along the $c-$ and $a-$axes, with a metal-to-insulator transition T$\rm_{MI}$ near 135~K \cite{YamadaJPSJ99, YamauchiSSS05}, which is very sensitive to the stoichiometry and imperfections \cite{YamadaJPSJ99}. The temperature-dependent magnetic susceptibility indicates a Curie-Weiss behavior accompanied with an antiferromagnetic ordering near 20~K (attributed to a weak coupling perpendicular to the $b-$axis), and an anomaly near T$\rm_{CO}$/T$\rm_{MI}$ \cite{SchlenkerJAP79}. 

At room temperature, the crystal structure of $\beta$-Na$_{0.33}$V$_2$O$_{5}$ is monoclinic (space group: C2/m) having six formula units per unit cell with a total of 44 atoms. This structure is highly anisotropic with three inequivalent sites of V-ions namely V1, V2, and V3, such that along the $b-$axis the V1O$_6$ octahedra form an edge-sharing zigzag chain. The V2 site has a similar octahedral coordination and forms a corner-sharing ladder chain and the V3 site is in fivefold square pyramidal coordination and arranged in an edge-sharing zigzag chain.  Moreover, the sodium ions occupy the empty space formed by the V$_2$O$_5$ framework. Galy \textit{et al.} reported that there are two different sites for the Na ions, M1 and M2, which have a separation of 1.95~\AA~in the $ac-$plane \cite{GalyJSSC70}. This distance does not allow simultaneous occupation of both sites (M1 and M2) because of the large ionic radii of Na$^+$ ions (1.02~\AA) \cite{GalyJSSC70,ShannonAC69}, while another report suggests that Na atoms can form a zigzag chain along the $b-$axis with a periodicity of 2$b$ \cite{MaruyamaJPSJ80}. Moreover, according to Goodenough, the V$^{4+}$ ions occupy 50\% V1 sites and a few of V3 sites \cite{GoodenoughJSSC70}, while a study done by Badot \textit{et al.} using the bipolaron model depicts that V$^{4+}$ ions are distributed between V1 and V3 sites in a ratio of 84\% and 16\%, respectively \cite{BadotJSSC91}. It was earlier reported from NMR measurements that monovalent sodium in this compound donates its outer shell electrons to the 3$d$-orbital of V-ions \cite{ItohPB00}. Goodenough has reported that these Na electrons overlap with the d$_{xz}$ orbital of the V1 sites or $d_{yz}$-orbital of the V2 sites, including a possibility of hopping into the V3 sites via the intermediate O atom \cite{GoodenoughJSSC70}. Interestingly, angle-resolved photoemission measurements revealed a strong electron-phonon coupling, and no significant change in the electronic properties across T$\rm_{MI}$ \cite{OkazakiPRB04}. Moreover, Raman spectroscopy has been utilized as a powerful tool to understand the lattice dynamics and distinguish the precise changes over the phase transitions in the vanadium bronzes \cite{PopovicJPCM03, PopovicJPCM06} as well as in other complex oxides \cite{AjayPRB20, DuaPRB21}. In this line, there are a few reports on the lattice dynamics and optical properties of the single-crystalline \cite{FrankPRB07, PresuraPRL03} and polycrystalline \cite{HadjeanJMC11, OsmanACSAMI21, PhuocPRB05} $\beta$-Na$_{0.33}$V$_2$O$_5$ samples. However, to the best of our knowledge, no detailed temperature dependence of the Raman spectrum has been performed yet. Therefore, it is vital to extensively investigate the physical properties of polycrystalline $\beta$-Na$_{0.33}$V$_2$O$_5$ with temperature. 

In this paper, we study the structural, magnetic, transport, electronic, and temperature-dependent vibrational properties of $\beta$-Na$_{0.33}$V$_2$O$_{5}$. The Rietveld refinement of the XRD pattern using monoclinic space group C2/m confirms the phase purity. The experimentally derived value of effective magnetic moment (0.63~$\rm\mu$B) manifests that the V-ions exist in the mixed-valence states of 4+ and 5+, which is further confirmed by x-ray photoemission spectroscopy and near-edge x-ray absorption spectroscopy at room temperature. Further, the curve fitting analysis of extended x-ray absorption fine structure (EXAFS) at the V K-edge gives an insight into the local structure. More importantly, the temperature-dependent Raman scattering probe the precise structural changes and stability of the compound with a dominant spin-phonon coupling at low temperatures. The dc-resistivity measurements reveal that system does not follow a single conduction mechanism in the measured temperature range and exhibits a semiconducting behavior.

\section{\noindent ~Experimental}

A single-phase $\beta$-Na$_{0.33}$V$_2$O$_{5}$ (NVO) sample was synthesized by the sol-gel route \cite{SarohaAO19}. The powder x-ray diffraction pattern was recorded using the Cu-K$\alpha$ radiation (1.5406~\AA) with a Panalytical x-ray diffractometer and Rietveld refinement was performed with the FullProf software using a linear background. The temperature dependence of the Raman spectra was recorded using a high-resolution Dilor XY triple-monochromator spectrometer in backscattering geometry  and the final signal was detected by a liquid nitrogen cooled charge-coupled-device, a CCD-camera. For excitation, the laser lines $\lambda=$ 632.82~nm of  He+/Ne+ mixed gas laser and $\lambda=$ 514.532~nm of an Ar+ laser were used. In order to avoid sample heating through the laser beam, the incident laser power was kept below 5~mW at the sample position with a diameter of the laser spot of about 0.1~mm. For measurements at low temperature, the samples were mounted in a helium-flow cryostat and for the high temperature measurements in an optical temperature control stage of the company Renishaw was used. For each Raman spectrum, an additional calibration spectrum of a nearby neon line was measured in order to determine the precise frequency and line-width of different phonons. The phonon modes were analyzed by fitting the Voigt profile to the experimental data, which corresponds to the deconvolution of the Lorentzian (phonon line) and Gaussian (instrumental resolution of $\sim$2~cm$^{-1}$ determined from neon spectrum) line shapes \cite{RahlenbeckPRB09}. The dc-magnetic susceptibility and dc-transport measurements were performed with the MPMS and PPMS instruments, respectively, from Quantum Design, USA. X-ray photoemission spectra were recorded using a commercial electron analyzer PHOIBOS150 from the company Specs, GmbH, Germany, which utilizes a nonmonochromatic Al-K$\alpha$ source (h$\nu=$1486.6~eV). The base pressure in the analysis chamber was of the order of 10$^{-10}$~mbar and the pass energy of the analyzer was 40~eV in transmission mode. The x-ray absorption spectroscopy (XAS) measurements were carried out in the transmission mode at the energy scanning EXAFS beamline (BL-09) of the INDUS-2 synchrotron source (2.5~GeV, 200~mA) at Raja Ramanna Center of Advanced Technology (RRCAT), Indore, India \cite{BasuJPCS14}. 

\section{\noindent ~Results and Discussion}

\begin{figure}[h]
\centering
\includegraphics[width=3.5in]{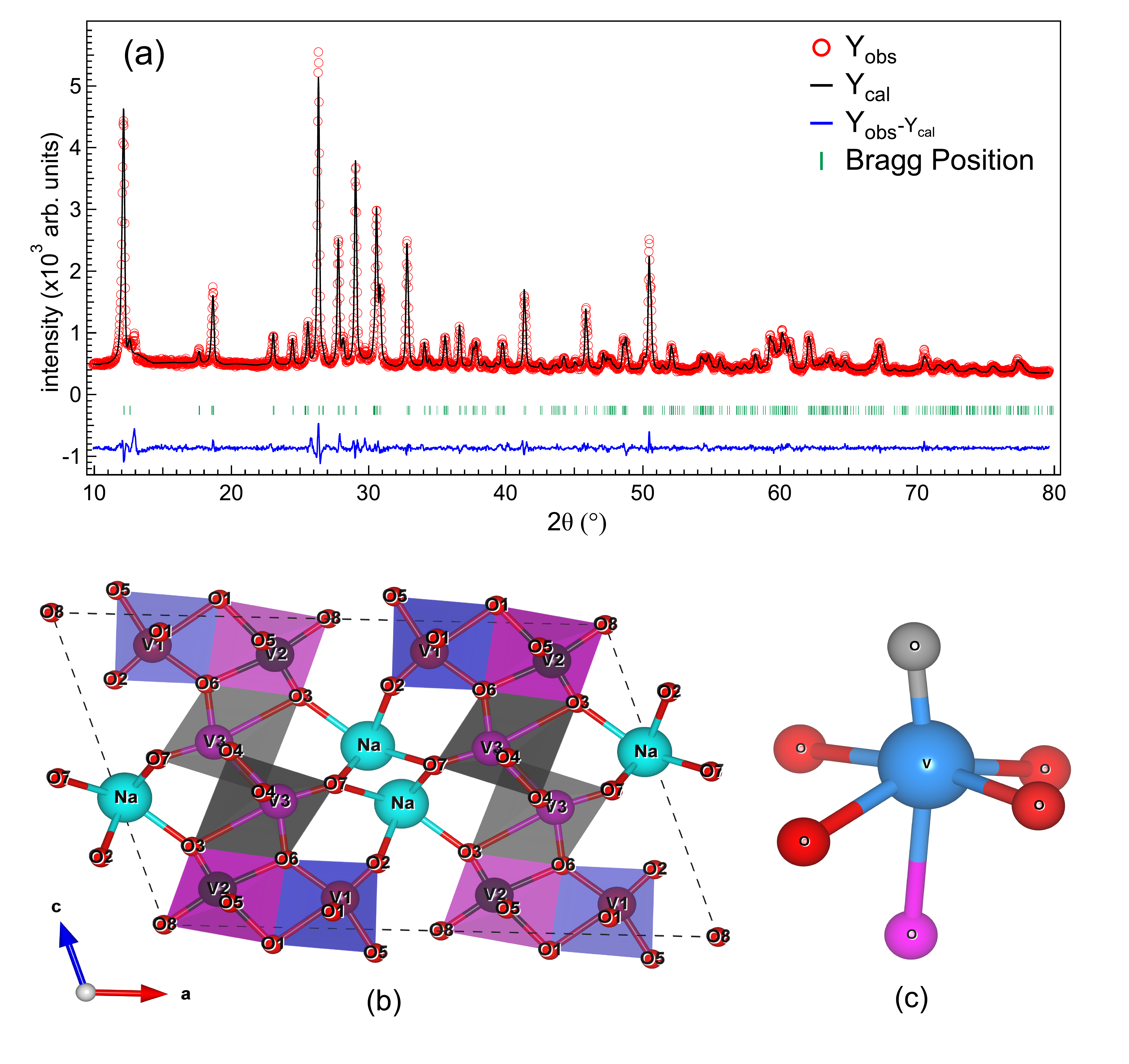}
\caption{(a) The Rietveld refinement of the XRD pattern of $\beta-$Na$_{0.33}$V$_2$O$_{5}$ sample. (b) A schematic of the crystal structure projected in the $ac-$plane (cyan color for Na; blue, magenta, and black polyhedra are formed around the V1, V2, and V3 sites, at different symmetry positions, and the red color represents the oxygen atoms), (c) the VO$_6$ octahedron and distribution of neighboring oxygen atoms manifests three groups of bonds as shown by grey (apical), red (basal), and magenta (farthest) colors.} 
\label{fig:XRD}
\end{figure}

The Rietveld refinement of powder x-ray diffraction pattern of $\beta$-Na$_{0.33}$V$_2$O$_5$ measured at room temperature is shown in Fig.~\ref{fig:XRD}(a), which confirms a monoclinic space group C2/m with the lattice parameters, $a=$ 15.4431(7)~\AA, $b=$ 3.6147(2)~\AA, $c=$ 10.0848(3)~\AA, and $\beta=$ 109.546(2)$\rm^o$  \cite{SarohaAO19}. 
The refinement parameters $\chi^2$ = 1.65, R$_{wp}$ = 5.28\%, and R$_p$ = 3.95\% manifest a good quality fitting of the pure phase XRD pattern. The two-dimensional (2D) projection of the crystal structure of $\beta$-Na$_{0.33}$V$_2$O$_5$ is presented in Fig.~\ref{fig:XRD}(b) within the $ac-$plane, where $b-$axis points inward. We find that there are three different symmetry positions of the V-ions [V1, V2, and V3], as well as tunnel sites for the accommodation of Na-ions (in cyan color) in the lattice. The blue, magenta, and black color polyhedra are formed around the V1, V2, and V3 sites, respectively, and the symmetry positions for the eight oxygen atoms (in red color) are also indicated with their respective numbers (O1--O8). Fig.~\ref{fig:XRD}(c) shows a schematic picture where there are three groups of V-O oxygen bonds having $\sim$1.6~\AA~for the apical oxygen (in grey color), $\sim$1.8--2.0~\AA~for the basal plane (in red color), and $\sim$2.3~\AA~for the farthest oxygen atom (in magenta color) \cite{DoubletPRB05, SmolinskiPRL98}. Figures~\ref{fig:XRD}(b, c) are constructed using VESTA 3 software for visualization of crystal structure \cite{MommaJAC11}. We present all the structural parameters including atomic positions and bond distances in Tables I and II of \cite{SI}, which are found to be in good agreement with reported in refs.~\cite{HadjeanJMC11, WadsleyAC55}. 

\begin{figure}[h]
\centering
\includegraphics[width=3.6in]{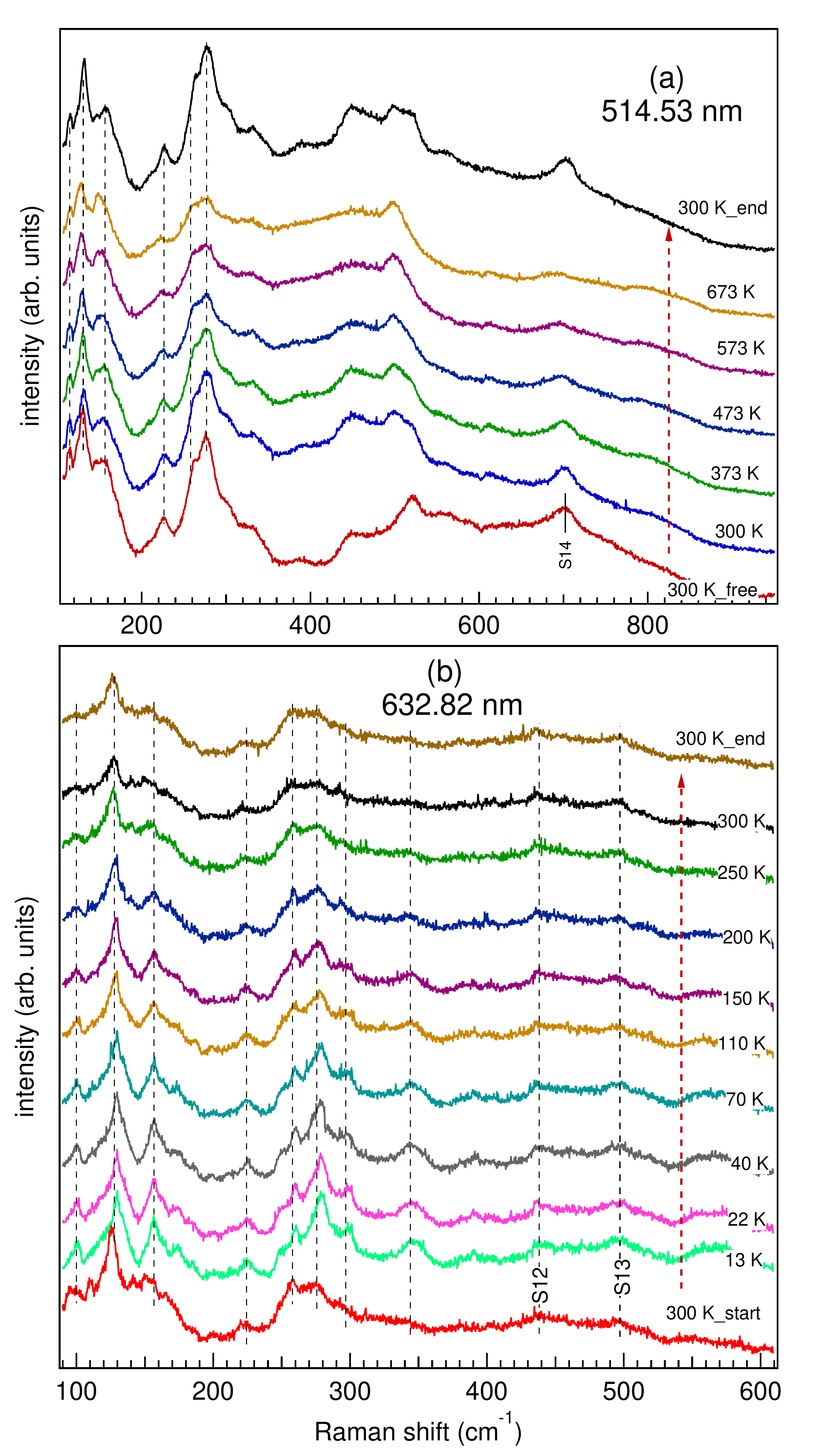}
\caption{The temperature-dependent Raman spectra of $\beta$-Na$_{0.33}$V$_2$O$_{5}$, measured with (a) 514.53~nm, and  (b) 632.82~nm laser wavelengths. The red dotted arrows in each graph represent the sequential measurement protocol and each curve is shifted vertically just for the clarity in presentation.} 
\label{fig:Raman}
\end{figure}

In order to understand the lattice dynamics in $\beta$-Na$_{0.33}$V$_2$O$_{5}$ sample, we have performed temperature-dependent Raman measurements in a large temperature range, i.e., from 13~K to room temperature (RT) using the wavelength $\lambda=$ 632.82~nm for excitation and from RT to 673~K using the $\lambda=$ 514.53 nm laser line. The spectra are presented in Figs.~\ref{fig:Raman}(a, b) between 107--950 cm$^{-1}$ and 90--610 cm$^{-1}$, respectively. Note that all the Raman spectra were calibrated with respect to a Neon line measured at each corresponding temperature and there was no significant variation in the recorded neon spectra with temperature. 
At room temperature, $\beta$-Na$_{0.33}$V$_2$O$_{5}$ has a monoclinic crystal structure of space group C2/m (point group: C$^3_{2h}$), which consists of six formula units (Z = 6) with 44 atoms per unit cell. Here, all the atoms have 4$i$ Wyckoff positions ($x$,0,$z$; and $-x$,0,$-z$) with site symmetry of C$^3_{2h}$ except for one oxygen (O1), which is located at the Wyckoff position 2$a$ (0,0,0) with C$_{2h}$ site symmetry \cite{PopovicJPCM03, WadsleyAC55}. The Na-ions are statistically distributed in the tunnel sites formed along the $b$-axis (not more than half-filled) by the atoms that occupy 4$i$ Wyckoff positions to maintain the stoichiometry of the compound \cite{WadsleyAC55}. According to Popovi\'c {\it et al.} \cite{PopovicJPCM03}, these symmetries result in a large number of optical modes, and according to the group theory, there are in total 66 phonon modes, which involve 3 acoustic (one A$_u$ and two B$_u$), 30 Raman, and 33 infra-red active modes. The factor group analysis yields the following distribution of vibrational modes \cite{PopovicJPCM03}:
\begin{equation}
\Gamma = 20A_g + 10B_g + 12A_u + 24B_u
\end{equation}
\begin{table}
 \caption{The experimentally observed frequencies ($\omega_{obs}$) of the Raman modes in $\beta$-Na$_{0.33}$V$_2$O$_{5}$ (NVO) at room temperature are compared with polycrystalline \cite{HadjeanJMC11}, and single-crystalline \cite{FrankPRB07} NVO samples. The assignment as well as origin of individual modes are summarized according to the published room temperature data on single-crystalline $A$V$_2$O$_5$ ($A=$ Na, Ca, Mg, Cs) \cite{PopovicPRB02} and NaV$_2$O$_5$ \cite{PopovicSSC99} samples.}  
 \vskip 0.15cm
 \centering 
 \begin{tabular}{|c|c|c|c|c|}
  \hline 
   &  & poly-crystal  & single-crystal  &  origin of \\ 
  modes & this work &  NVO &  NVO &  individual mode\\ 
  &$\omega_{obs}$ &$\omega_{exp}$ \cite{HadjeanJMC11}& $\omega_{exp}$ \cite{FrankPRB07} &\cite{PopovicPRB02, PopovicSSC99}\\[0.5ex]
  \hline
  S0& A$_g$(100) &-- & -- & chain rotation\\
  S1& A$_g$(115) &  -- & --& --\\
  S2& A$_g$(127.5)  & A$_g$(124) &--&  -- \\
  S3& B$_g$(152) &B$_{g}$(151)&B$_{g}$(154)&Na$\parallel$c\\
  S4& A$_g$(170) & -- & -- &\begin{tabular}{c}Na$\parallel$b\\Chain rotation\end{tabular}\\
  S5& A$_g$(222.5)  & A$_g$(223)& A$_g$(230) &\begin{tabular}{c}O$_8$-V$_3$-O$_7$\\bending\end{tabular} \\
  S6& A$_g$(249)   & --&--&--\\
  S7& B$_g$(257.5)  &A$_g$(256)& B$_g$(256)  &\begin{tabular}{c}O-V-O\\bending\end{tabular}\\
  S8& B$_g$(274)  & B$_g$(275)& B$_g$(275)&\begin{tabular}{c}O-V-O\\bending\end{tabular}\\
  S9& A$_g$(295)  & A$_g$(288)&A$_g$(298) &\begin{tabular}{c}O$_8$-V$_3$-O$_7$\\bending\end{tabular} \\
  S10& B$_g$(318) & -- & -- &--\\
  S11& A$_g$(343)  & A$_g$(333)  & A$_g$(331) &\begin{tabular}{c}O$_3$-V$_2$-O$_4$\\bending\end{tabular} \\
  S12& A$_g$(438)  & A$_g$(440)&A$_g$(440) &\begin{tabular}{c}V-O$_3$-V\\bending\end{tabular} \\
  S13& A$_g$(497) & A$_g$(516) & A$_g$(500) &\begin{tabular}{c}V$_3$-O$_{6}$\\stretching\end{tabular} \\
  S14& B$_g$(702.5)  & B$_g$(697)  & B$_g$(694) &\begin{tabular}{c}V$_3$-O$_{4}$\\stretching\end{tabular} \\
\hline
 \end{tabular}
 \label{Tab:Raman}
\end{table} 

As the Raman measurements were done without polarization analysis, both the A$_g$ and B$_g$ modes are present in the spectrum. The mode assignment was performed by comparing the observed phonon frequencies (listed in Table~I) with the data from refs.~\cite{FrankPRB07, HadjeanJMC11}. The phonon modes in the range of 200--500 cm$^{-1}$ originate from the bond bending vibrations, whereas the phonon modes at higher wavenumber ($>$ 500 cm$^{-1}$) originate from the stretching vibrations of different V--O bonds of the VO$_{5/6}$ polyhedra in the structure having different symmetry environments. In general, the modes originating from the polyhedra are collective vibrations of the oxygen ions at their corners, and the vibrational modes originating from the octahedra appear at lower wavelengths than the modes of pyramidal units due to the larger reduced mass of the contributing oxygen ions \cite{LoaPSS99}. Doublet \textit{et al.} reported that five out of the six V-ions in the $\beta$-phase compounds are surrounded by the five nearest oxygen neighbors to form the distorted square pyramid, while the sixth oxygen atom is found at a higher distance as compared to the other five oxygen atoms in the VO$_6$ octahedron \cite{DoubletPRB05}, as shown in Figs.~\ref{fig:XRD}(b, c). Since there are three inequivalent V-sites available in $\beta$-Na$_{0.33}$V$_2$O$_{5}$, a weighted average of the V-O bond lengths have the values 1$\times$1.588~\AA~(apical); 4$\times$1.901~\AA~(basal); and 1$\times$2.315~\AA~(farthest) in three groups (obtained from the analysis of EXAFS spectra, see Table III for details) \cite{PopovicJPCM03, DoubletPRB05, FrankPRB07}. Moreover, a few weaker Raman modes are present at higher wavenumbers; for example, at 438, 497, and 702.5~cm$^{-1}$, as marked by S12, S13, and S14, respectively, in Fig.~\ref{fig:Raman}. Here, the S12 mode is assigned to A$_g$ vibrations  \cite{PopovicJPCM06, PopovicJPCM03, FrankPRB07, HadjeanJMC11}, and the peak position (which is sensitive to the Na concentration \cite{BacsaPRB00}) indicates that the electrons in the polycrystalline $\beta$-Na$_{0.33}$V$_2$O$_{5}$ are delocalized to the O-V-O ladder \cite{PopovicJPCM03}, where O refers to the bridge O-atoms between the VO$_{5/6}$ polyhedra. Therefore, the responsible bond for the S12 mode can be V$_3$-O$_5$-V$_1$ (134.935$^o$, 3.716~\AA), as also reported in Ref.~\cite{PopovicPRB02}. Intriguingly, the delocalization of charge in the V$_3$-O$_5$-V$_1$ ladder exhibit that V$_1$ and V$_3$ sites are mixed valent (4+ and 5+), while the V$_2$ site will be in 5+ state \cite{BacsaPRB00}. The S13 and S14 modes are associated to A$_g$ and B$_g$ vibrations due to the V$_3$-O$_6$ (1.919~\AA) and V$_3$-O$_4$ (1.881~\AA) bond stretching, respectively, which are consistent with $\alpha'$-NaV$_2$O$_{5}$, as reported in \cite{PopovicPRB02, PopovicSSC99}. In Table~I, we compare the peak positions of different Raman modes and their respective origins with the help of Refs.~\cite{HadjeanJMC11, FrankPRB07, PopovicPRB02, PopovicSSC99}. 

\begin{figure}[h]
\centering
\includegraphics[width=3.4in]{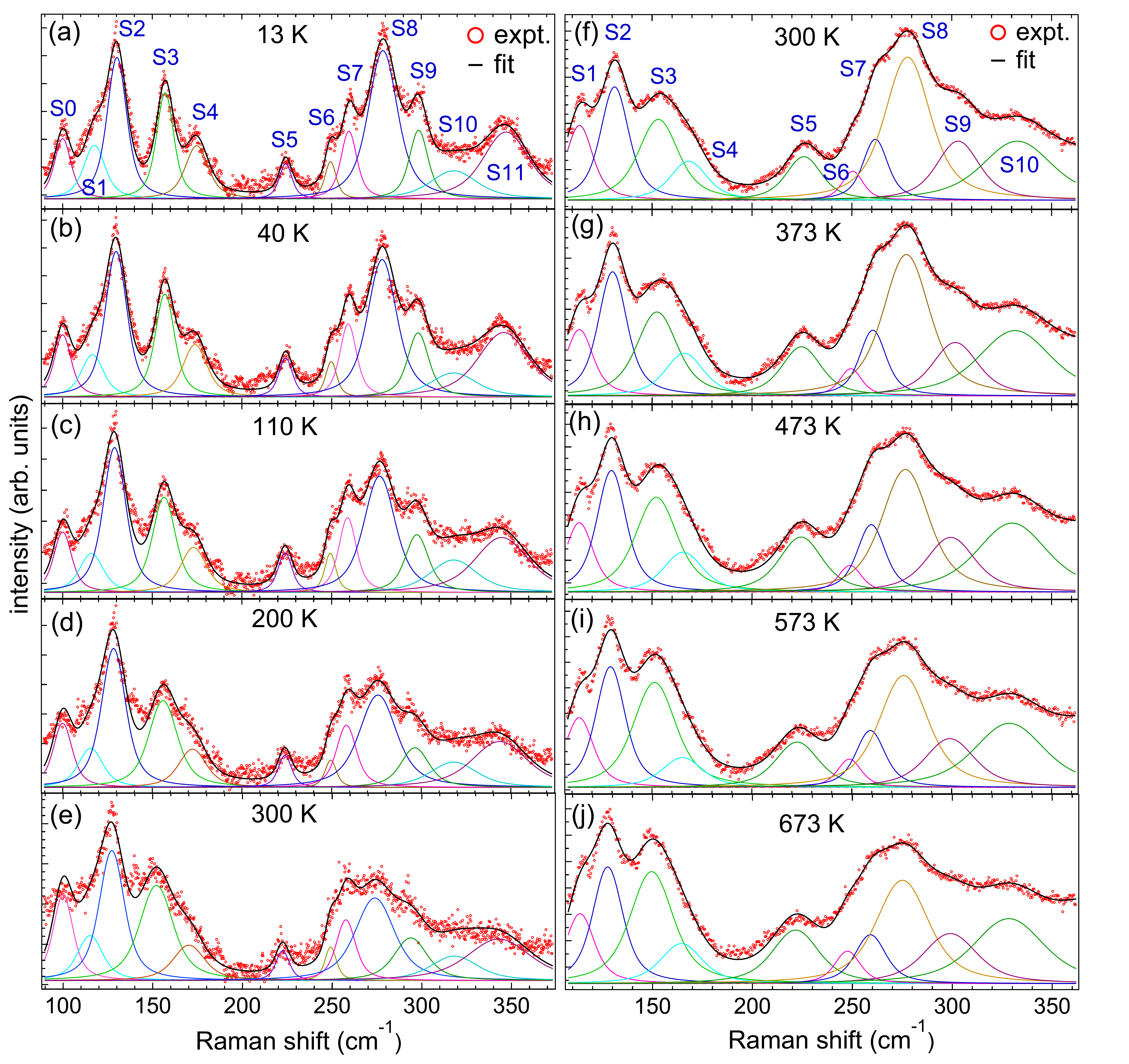}
\caption{Raman spectrum of $\beta$-Na$_{0.33}$V$_2$O$_{5}$ at selected temperatures measured with (a-e) 632.82~nm, and (f-j) 514.53~nm excitation wavelengths. The solid lines present the fitted individual modes using the Voigt function and a solid thick black line presents the total fit. All the Raman modes are marked with the S0--S11 notation according to their position and their assignment is presented in Table~I.}
\label{fig:Raman_Fit}
\end{figure}

Moreover, we analyze the recorded temperature-dependent Raman spectra below 400~cm$^{-1}$ by fitting the Voigt peak profiles (marked as S0--S11), as presented in Figs.~\ref{fig:Raman_Fit}(a--j) for selected temperatures \cite{RahlenbeckPRB09, BuchenauPRB20, GrahamPRB22}. The Raman modes with the frequencies $\le$200 cm$^{-1}$ (S0--S4) are either associated with the chain rotation/translation or with the movement of Na-atoms along the different crystallographic axis. As the unit cell parameters follow $a>c>b$ \cite{MeetsmaACC98}; therefore, the smaller, mid, and high-frequency modes in $\beta$-Na$_{0.33}$V$_2$O$_{5}$ are predominantly associated with the movement of sodium atoms along the longer to the smaller axis, respectively, i.e., along the $a$, $c$, and $b-$axes, respectively. Moreover, the phenomenon of chain rotation/translation associated with the S0 and S4 Raman modes manifest that these modes correspond to the motion of constituent chains, where the movement of all atoms (V, Na, and O) takes place in a systematic pattern. While the Raman modes which appear due to the translation of Na atoms (S3, S4), the only motion of Na atoms takes place along a defined crystallographic axis. Further, S1 and S2 Raman modes belong to the A$_g$ type and will have a similar origin as S3 and S4 modes. Interestingly, the position of the remaining seven active Raman modes (S5--S11) lies in the frequency range of bond-bending vibrations, in which S5, S6, S9, and S11 modes are of A$_g$ type, while S7, S8, and S10 modes are related to the B$_g$ type vibrations [see Table-I]. The S5 (222.5~cm$^{-1}$) and S9 (295~cm$^{-1}$) modes are analogous to the A$_g$ vibrations observed in the $aa$ and $bb$ polarizations, respectively, which manifest the different types of bond bending vibrations of O-V-O bonds \cite{FrankPRB07}. Note that in $\alpha'$-NaV$_2$O$_5$, the Raman modes at 233 and 304~cm$^{-1}$ are related to the O$_1$-V-O$_2$ bond bending vibrations having the bond lengths and bond-angles of 3.552~\AA, 108.488$^{\rm o}$; and 3.632~\AA, 108.428$^{\rm o}$, respectively \cite{PopovicPRB02}. Therefore, the A$_g$ type S5 and S9 vibrational modes in $\beta$-Na$_{0.33}$V$_2$O$_5$ are associated with the O$_8$-V$_3$-O$_7$ bond having the dimensions of 3.712~\AA~ and 112.752$^{\rm o}$, which are very similar to the $\alpha'$-NaV$_2$O$_5$. Moreover, the S11 mode at 343~cm$^{-1}$ is also of A$_g$ type, but is associated with the O$_3$-V$_2$-O$_4$ bond (3.437~\AA, 105.364$^{\rm o}$), which was found to be very similar to the O$_3$-V-O$_1$ bond bending (3.442~\AA, 103.455$^{\rm o}$) in $\alpha'$-NaV$_2$O$_5$ \cite{PopovicPRB02}. The S7, S8, and S10 modes are of B$_g$ type observed at 257.5, 274, and 318~cm$^{-1}$, respectively, and correspond to the O-V-O bond bending vibrations. 

\begin{figure}
\centering
\includegraphics[width=3.5in]{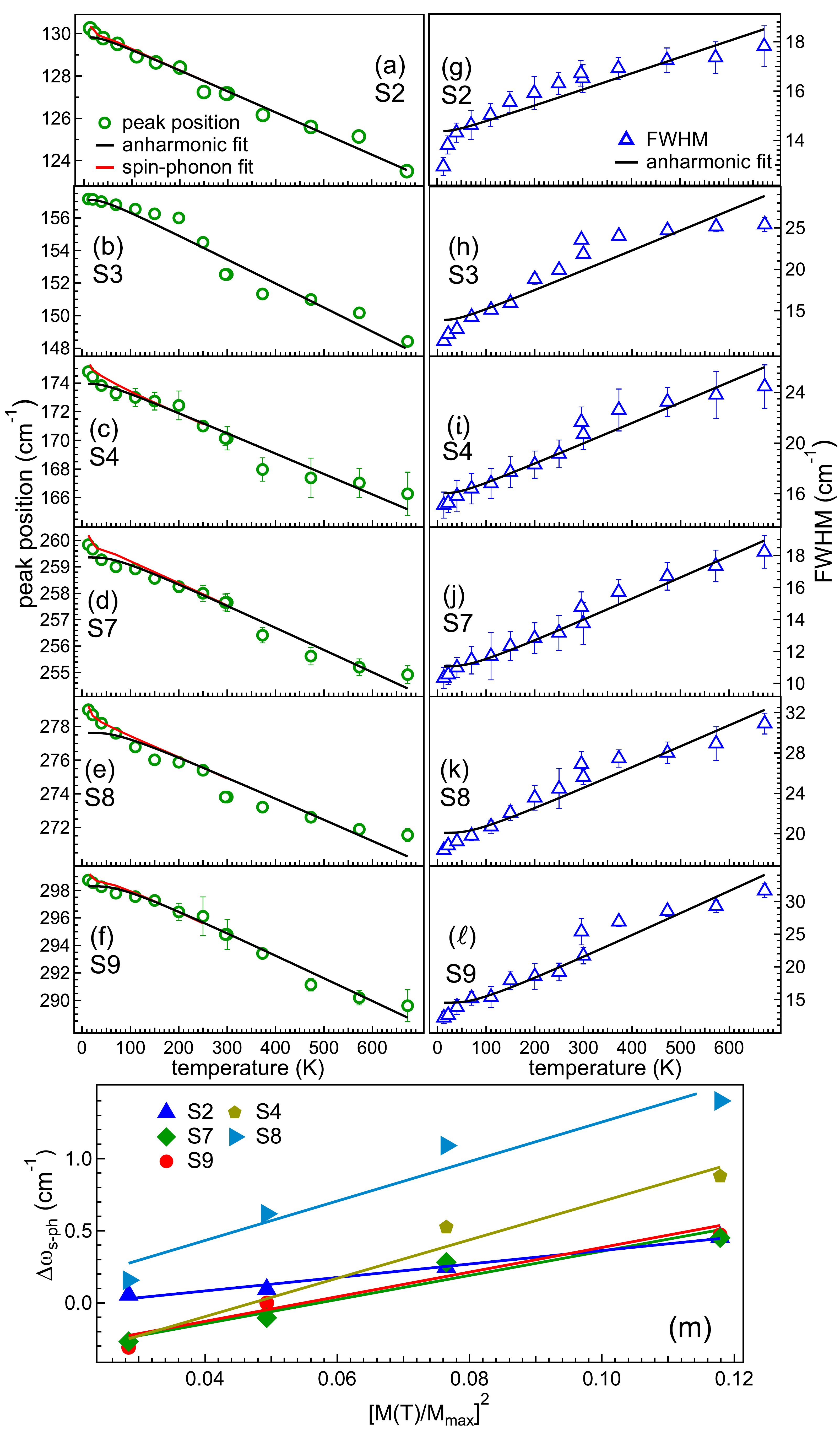}
\caption{The temperature dependence of the (a--f) phonon frequency and (g--l) full width half maximum (FWHM) of the Stokes lines (S2, S3, S4, S7, S8, and S9) in the range of 13--673~K. The green open circles and blue open triangles represent the experimentally determined values of the peak positions and FWHM, respectively. The solid black lines are the fit of an anharmonic phonon decay using equations (2) and (3), and the solid red lines represent the fit corresponding to the spin-phonon interaction using equations (4) and (5). (m) The plot between $\Delta\omega_{s-ph}$ and [M(T)/M$_{max}$]$^2$ at $\le$70~K for selected Raman modes and solid lines show the linear fit.}
\label{fig:FWHM+Position}
\end{figure}

Figures~\ref{fig:FWHM+Position}(a--f) and \ref{fig:FWHM+Position}(g--l) show the peak positions (open green circles) and full width at half maximum (FWHM) (open blue triangles), respectively, of only high intensity and low wavenumber Raman modes, i.e., S2 (127.5~cm$^{-1}$), S3 (152~cm$^{-1}$), S4 (170~cm$^{-1}$), S7 (257.5~cm$^{-1}$), S8 (274~cm$^{-1}$), and S9 (295~cm$^{-1}$) as a function of temperature. We find that all these modes shift towards lower wavenumber and the FWHM increases with temperature, as plotted on the left and right axis of Figs.~\ref{fig:FWHM+Position}(a--f) and \ref{fig:FWHM+Position}(g--l), respectively. Since the temperature predominantly alters the equilibrium positions of atoms in the lattice owing to the changes in the interatomic forces due to the lattice thermal expansion and anharmonicity. These effects can be considered as composed of two different contributions: (i) change in the frequency of phonos originating because of volume expansion with temperature, which is the 'implicit' anharmonicity, and (ii) the pure-temperature contribution due to anharmonic phonon-phonon interactions, called 'explicit' anharmonic contribution \cite{LucaJRS03, MaczkaJPCM12}. Therefore, the change in phonon frequency and linewidth with temperature is caused by the anharmonic decay of the optical phonon into two phonons of lower energy. This can be expressed in terms of a phonon self-energy, where the real part corresponds to the phonon frequency shift and the imaginary part is related to the change in the FWHM \cite{BalkanskiPRB83, KlemensPR66, MenendezPRB84}. This can be modeled in first approximation by a symmetric decay (Klemens decay) into two phonons with half the energy of the optical phonon and opposite momenta, and can be expressed as \cite{MenendezPRB84, RahlenbeckPRB09}:
\begin{equation}
\omega_{anh}(T) = \omega_0 + A\left(1+\frac{2}{exp(\hbar\omega_0/2k_BT)-1}\right)
\end{equation}
where A is a constant and {$\omega_0$ is the frequency of the corresponding optical phonon at lowest temperature. Similarly, we can assume this for the phonon line widths of Raman modes using a positive constant B, and the line width at lowest temperature ($\Gamma_0$), such that 
\begin{equation}
\Gamma_{anh}(T) = \Gamma_0 + B\left(1+\frac{2}{exp(\hbar\omega_0/2k_BT)-1}\right)
\end{equation} 
The lifetime of phonon decay can be determined using the energy-time uncertainty relation $\tau = \hslash/\Delta E = 1/(2\pi{c}\Gamma)$, where $\tau$ is lifetime of phonon mode, $\Delta E$ is uncertainty in the phonon energy, $\hslash$ is reduced Planck's constant, $\Gamma$ is phonon linewidth and $c$ is speed of light \cite{BergmanPRB99}. Since the linewidth increases with temperature, the lifetime of the phonon decreases [right axis of Figs.~\ref{fig:FWHM+Position}(g--l)]. The obtained values of $\tau$ (at lowest temperature) are 0.37$\pm$0.02~ps, 0.46$\pm$0.05~ps, 0.35$\pm$0.02~ps, 0.54$\pm$0.03~ps, 0.29$\pm$0.02~ps, and 0.48$\pm$0.08~ps for S2, S3, S4, S7, S8, and S9 Raman modes, respectively, consistent with  Refs.~\cite{DeepaMTC21, BergmanPRB99}. The anharmonic model was used to describe the temperature dependence of the phonon frequency and FWHM. Therefore, we use equations (2) and (3) to fit the temperature dependence of the phonon frequency and FWHM, respectively. The corresponding results are shown as the solid black lines in Figs.~\ref{fig:FWHM+Position}(a--l) and the obtained fitting parameters at lowest temperature are summarised in Table-II. 

\begin{table}
 \caption{Anharmonic fit parameters determined for the phonon frequency ($\omega_0$, A) and full-width at half-maximum ($\Gamma_0$, B) of the individual Raman modes of $\beta$-Na$_{0.33}$V$_2$O$_{5}$.}
 \vskip 0.15cm
 \centering 
 \begin{tabular}{|c|c|c|c|c|}
  \hline 
  ~Raman~& $\omega_0$ & A & $\Gamma_0$ & B\\ 
  mode& (cm$^{-1})$ & (cm$^{-1}$) & (cm$^{-1}$) & (cm$^{-1}$)\\[0.5ex]
  \hline
    S2 &~~129.8$\pm$0.1~~&~~-0.47$\pm$0.02~~&~~14.1$\pm$0.3~~&~~0.31$\pm$0.04~~\\
  S3&~157.1$\pm$0.4~&~-0.84$\pm$0.07~&~12.6$\pm$0.9~&~1.36$\pm$0.16~\\
  S4&~174.0$\pm$0.3~&~-0.90$\pm$0.06~&~15.1$\pm$0.4~&~1.02$\pm$0.08~\\
  S7&~259.4$\pm$0.2~&~-0.79$\pm$0.04~&~9.8$\pm$0.3~&~1.26$\pm$0.07~\\
  S8&~277.6$\pm$0.4~&~-1.27$\pm$0.13~&~17.9$\pm$0.6~&~2.11$\pm$0.19~\\
  S9&~298.3$\pm$0.2~&~-1.81$\pm$0.07~&~10.9$\pm$0.9~&~3.37$\pm$0.31~\\
  \hline
 \end{tabular}
 \label{Tab:Anharmonic_Fit}
\end{table} 

It is interestingly to note here that the measured data for the peak positions and linewidth deviate from the fitted line of anharmonic phonon-phonon interactions in the low-temperature range below 40~K [see Figs.~\ref{fig:FWHM+Position}(a--l)]. This deviation can be attributed to an additional spin-phonon interaction of the Raman modes, as the magnetization data also show a rapid increase in the low temperature range [see Fig.~\ref{fig:MH+RT}(a), discussed later]. Here, the change in the phonon frequency due to the spin-phonon interaction can be calculated through a Taylor expansion of the exchange interaction with respect to the phonon displacement following the Refs.~\cite{BaltenspergerHPA68, LockwoodJAP88, GranadoPRB99}. A relation between the phonon frequency and spin-correlation function is given by:
\begin{equation}
\omega (T) = \omega_{anh}(T) + \lambda <S_i.S_j>
\end{equation}
where $\omega_{anh}$ is the anharmonic phonon frequency, $\lambda$ is the spin-phonon coupling constant and $<S_i.S_j>$ is the nearest neighbor spin-correlation function. Therefore, deviation in the phonon frequency beyond anharmonicity correspond to the spin-phonon interaction following the formalism developed in Ref.~\cite{GranadoPRB99}. It can be written in terms of the temperature-dependent magnetization of the magnetic Vanadium (V) \cite{GranadoPRB99}:
\begin{equation}
\Delta\omega_{s-ph} (T) =  \lambda <S_i.S_j> = n\lambda \left[\frac{M(T)}{M_{max}}\right]^2
\end{equation}
where $n$ is the number of interacting nearest neighbors in a plane. Here, the values of the temperature-dependent magnetization M(T), and the maximum value of the magnetization M$_{max}$ were determined independently from the ZFC data. In Fig.~4(m), we have plotted the deviation in the phonon-frequency from beyond the anharmonic phono-phonon interaction with [M(T)/M$_{max}$]$^2$ in the temperature range below $\approx$70~K. By a linear fit to these experimental data points using equation (5), we are able to determine the spin-phonon coupling constant ($\lambda$) for the individual phonons and the obtained values in cm$^{-1}$ are 1.2$\pm$0.1, 3.3$\pm$0.4, 2.1$\pm$0.4, 3.4$\pm$0.6, and 2.2$\pm$0.4 for S2, S4, S7, S8, and S9 Raman modes, respectively. These values are consistent with the one reported for the ZnCr$_2$O$_4$ (3.2--6.2~cm$^{-1}) $\cite{SushkovPRL05}. The obtained values of $\lambda$ manifest the strength of the spin-phonon coupling and the highest value of $\lambda$ for the S8 Raman mode exhibits that this mode is very sensitive to the spin-phonon coupling and this bending mode would be associated with the magnetic V ions (V1 and/or V3 sites) in the material. However, a similar value of $\lambda$ (3.3~cm$^{-1}$) is obtained for the S4 mode, which is associated with the translation and rotational motion of atoms and affected by the spin fluctuations. The lowest value of the $\lambda$ (1.2~cm$^{-1}$) for the S2 mode manifests a lower sensitivity to the spin-phonon coupling. Moreover, in Figs.~\ref{fig:FWHM+Position}(a--f), we have included an additional fit to the temperature-dependent peak position of the Raman modes corresponding to the spin-phonon interaction with a solid red line, which manifests a good agreement to the experimental data. Therefore, spin-phonon interaction in the low-temperature region is dominant and leads to the hardening of the phonon frequency. As shown in Figs.~\ref{fig:FWHM+Position}(g-l), the line width (FWHM) also decreases below about 40~K. The phonon self-energy consists of a real and an imaginary part, where the real part corresponds to the shift in phonon frequency caused by the spin-phonon interaction and the imaginary part is reflected in the change in the line width. This is a further support that the additional changes in phonon frequency and line width below 40~K are caused by the spin-phonon interaction and not effects such as lattice distortions \cite{BuchenauPRB20, GrahamPRB22}.

Now we move to the investigation of magnetic and transport properties of the $\beta$-Na$_{0.33}$V$_2$O$_{5}$ sample, as presented in Fig.~\ref{fig:MH+RT}. The temperature-dependent zero-field cooled (ZFC) and field cooled (FC) dc-magnetic susceptibility data are shown in Fig.~\ref{fig:MH+RT}(a) in the temperature range of 5--320 K. The ZFC-FC curves exhibit a paramagnetic type behavior where the magnetization value rises to 25~emu/mol at 5~K. Note that a hump near 45~K is due to measurement artifacts, i.e., attributed to the oxygen freezing on the sample surface \cite{KoblerPRB87}. The Curie-Weiss fit of the $\chi^{-1}$ versus temperature plot in 150 to 320~K temperature range results in an effective magnetic moment value of $\mu_{\rm eff}$ = 0.63~$\mu_{B}$ and a negative Curie-Weiss temperature $\theta_{\rm CW}$ = -52~K. The negative value of $\theta_{\rm CW}$ indicates the presence of antiferromagnetic spin correlations in the sample \cite{IsobeJPSJ96, VasilevPRB01}. Moreover, in the spin only approximation, we can determine the value of S$_{av}$ from the experimental results using the formula $\mu_{\rm eff} = 2\sqrt{S(S+1)}$, which gives the value of S$_{av}$ = 0.09. The V$^{3+}$ has two unpaired electrons, V$^{4+}$ has only one unpaired electron and V$^{5+}$ has no unpaired electron in the outer shell. Therefore, in the spin-only approximation, V$^{3+}$ and V$^{4+}$ will possess the $\mu_{\rm eff}$ values of 3.46~$\mu_{B}$ and 1.73~$\mu_{B}$, respectively, while V$^{5+}$ will give no contribution. Interestingly, in the unit cell of $\beta$-Na$_{0.33}$V$_2$O$_{5}$, the V-ions occur in 4+ and 5+ valence states in a ratio of 1 and 5, respectively \cite{HeinrichPRL04}. This will give the calculated value of $\mu_{\rm eff}=$ 0.61~$\mu_{\rm B}$, which is in good agreement with the experimentally obtained value of 0.63~$\mu_{B}$; therefore, the experimentally obtained share of V$^{4+}$ ions (18\%) in $\beta$-Na$_{0.33}$V$_2$O$_5$ is very close to the stoichiometric (16.7\%) value and also in a good agreement with the previous reports \cite{HeinrichPRL04}. The isothermal magnetization loop recorded at 5~K with an applied magnetic field of $\pm$70~kOe is shown in Fig. \ref{fig:MH+RT}(b)], which exhibits a paramagnetic type behavior for $>\pm$20~kOe, and an S-shaped below $\pm$20~kOe. This manifests a weak ferromagnetic behavior. The maximum value of magnetization is observed $\sim$125~emu/mol at $\pm$70~kOe with the coercivity and retentivity values of 175~Oe and 0.75~emu/mol, respectively. Further, we measure the virgin curve at 5~K and this data is utilized to get Arrott's plot in terms of M$^2$ versus H/M, as shown in Fig.~\ref{fig:MH+RT}(c). It is well known that a linear fit in higher magnetic field region indicates ferromagnetic interaction if it has an intercept on the y-axis and suggests a non-ferromagnetic interaction if the intercept is on the x-axis. In the present case, the extrapolation of the fit [black solid line in Fig.~\ref{fig:MH+RT}(c)] gives an intercept on the x-axis, which indicates a non-ferromagnetic interaction in the $\beta$-Na$_{0.33}$V$_2$O$_{5}$ sample. This is consistent with non-saturating nature of the M-H loop in Fig.~\ref{fig:MH+RT}(b). 

\begin{figure}[h]
\centering
\includegraphics[width=3.6in]{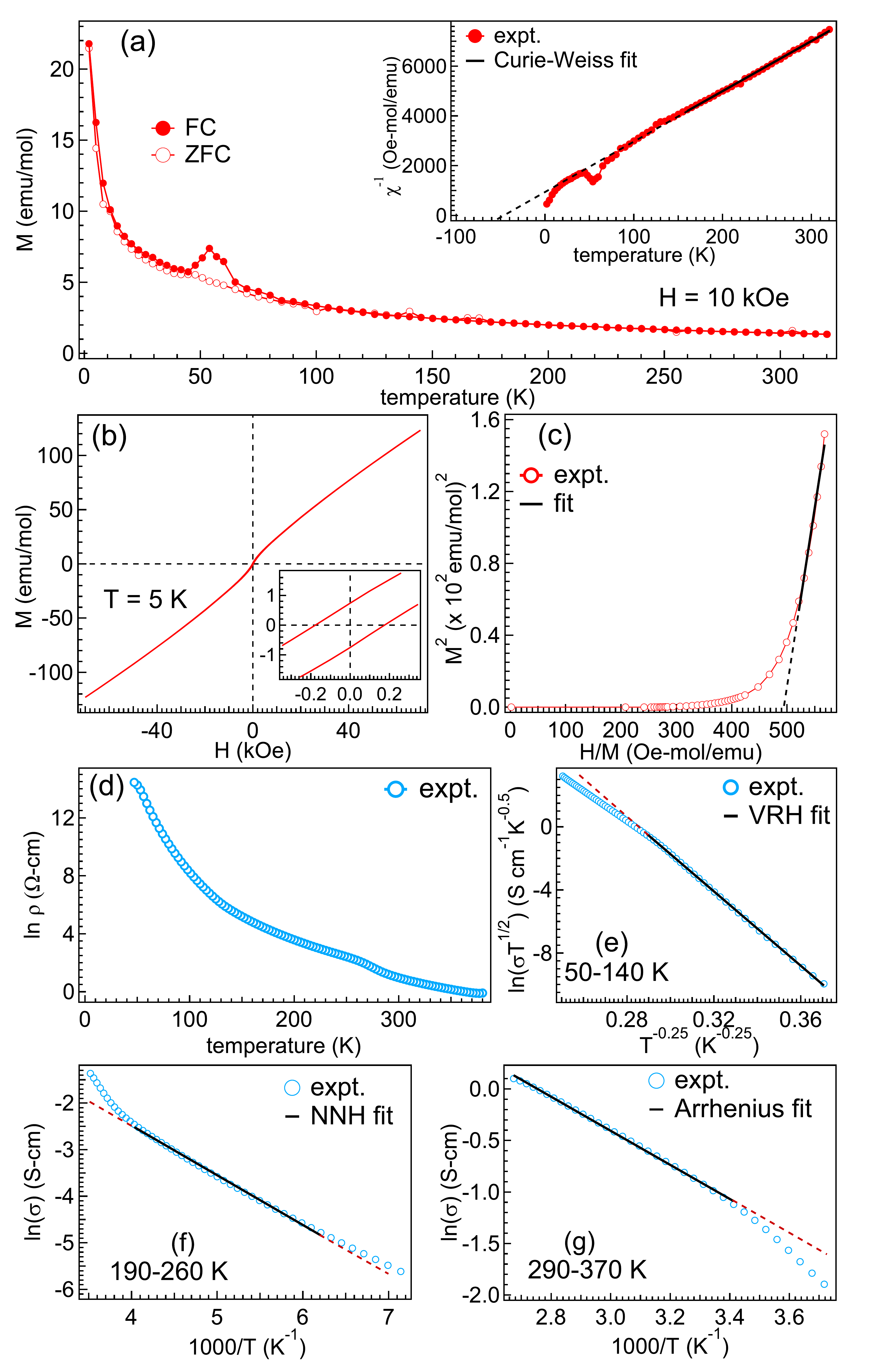}
\caption{(a) Temperature-dependent ZFC-FC curves of $\beta$-Na$_{0.33}$V$_2$O$_{5}$ with an applied field of 10~kOe, inset shows the Curie-Weiss fit of the FC data, (b) isothermal M-H loop recorded at 5~K, and the magnetic behavior near the origin is presented in the inset, (c) the Arrott's plot (M$^2$ versus H/M) from the virgin data at 5~K, and (d) the temperature-dependent resistivity data, with the fitting for the three different ranges: (e) low temperature 50--140~K using the 3D-variable range hopping model, high-temperature range of (f) 260--190~K with the nearest--neighbor hopping model, and (g) 290--370~K using an Arrhenius model. The solid black line in (e--g) manifests the fitting of the experimental data and the dotted red lines exhibit the range where the experimental data deviate from the respective models.} 
\label{fig:MH+RT}
\end{figure}

It has been reported that the single-crystalline $\beta$-Na$_{0.33}$V$_2$O$_5$ is a quasi-one-dimensional conductor along the $b-$axis at high temperatures, while semiconducting along the $a$ and $c-$axis \cite{YamadaJPSJ99, ObermeierPRB02}. To investigate the conduction mechanisms/pathways in polycrystalline $\beta$-Na$_{0.33}$V$_2$O$_5$ we have measured the resistivity in the temperature range of 50--380~K, as shown in Fig.~\ref{fig:MH+RT}(d). We find that the resistivity increases from $\Omega\cdot$cm to M$\Omega\cdot$cm range with decreasing temperature and exhibits a semiconducting nature due to a negative temperature coefficient. In semiconductors, high-temperature carrier transport is dominated by the thermally activated band conduction, where the charge carriers are thermally excited from localized states to delocalized states. However, in the low-temperature regime hopping of charge carriers from localized state to conduction band dominates. Interestingly, the conduction via hopping of charge carriers has two different possibilities (i) nearest-neighbor hopping (NNH) mechanism where the charge carriers hop from one localized state to the nearest unoccupied state, and (ii) variable range hopping (VRH) model where the carrier transport is facilitated to the unoccupied states near to the Fermi level despite their spatial distribution. The universal law of conduction follows \cite{BougiatiotiJAP17, ShuklaPRB18} 
\begin{equation}
\sigma_0 = \sigma_0 exp\left[\left(\frac{E_t}{k_BT}\right)^P\right]
\end{equation}
where $\sigma_0$ is the pre-exponential factor, E$_t$ is the  thermal activation energy, and $P$ is the characteristic exponent. The value of $P$ is 1 for the band conduction and between 0 to 1 for the hopping type conduction. The law for the NNH and thermally activated band conduction is defined as 
\begin{equation}
\sigma_0 = \sigma_1 exp\left[\left(-\frac{E_{a1}}{k_BT}\right)\right]+\sigma_2 exp\left[\left(-\frac{E_{a2}}{k_BT}\right)\right]
\end{equation}
where $\sigma_1$ and $\sigma_2$ are the pre-exponential factors and E$_{a1}$ and E$_{a2}$ are the thermal activation energies for the band and NNH conductions, respectively. To understand these mechanisms, we have plotted $ln(\sigma)$ versus 1000/T in the temperature ranges of 190--260~K and 290--370~K in Figs.~\ref{fig:MH+RT}(f, g) where the solid black lines in each panel show the fit to the experimental data using NNH and Arrhenius models, respectively, and the red dotted lines exhibit the range where the experimental data deviates from the fitting. The obtained activation energy values are 92$\pm$0.8~meV and 145$\pm$1~meV, for the NNH and band conduction, respectively, and are consistent with other semiconducting oxide materials \cite{ZobelPRB02, ShuklaPRB18}. 

More interestingly, the Mott-VRH type carrier conduction mechanism dominates in the low-temperature range \cite{MottCP69}. A strong coupling between the V$^{5+}$ and V$^{4+}$ ions in this sample creates delocalized orbitals. Therefore, the smaller values of the activation energies for hopping via an intermediary anion may be anticipated \cite{GoodenoughJSSC70}. Thus, the hopping via an intermediary cation goes through the V1/V3 subarray. However, with this kind of conduction in the present sample, any mobile electron will be confined to the single-channel/tunnel, which results in the large anisotropy in the resistivity. The Mott-VRH model assumes that the density of states near the Fermi level is constant and Coulomb repulsion between electrons is very weak and can be neglected \cite{PaulPRL73}. Therefore, for the VRH model, the value of $P$ is 0.25 and E$_t$ is defined as k$_B$T$_0$, where T$_0$ is the characteristic temperature and corresponds to the Mott energy E$_0$. The law for 3D-VRH conduction is: 
\begin{equation}
\sigma = \sigma_{H} exp\left[-\left(\frac{T_0}{T}\right)^{-1/4}\right]
\end{equation}
where $\sigma_H$ is the pre-exponential factor, and T$_0$ is the characteristic temperature. The $\sigma_H$ and T$_0$ are related to the inverse localization length ($\alpha$) and the density of states near the Fermi level N(E$_F$) by the  relations: 
\begin{equation}
\sigma_H=\left(\frac{3e^2\nu_{ph}}{(8\pi)^{1/2}}\right)\left(\frac{N(E_F)}{\alpha k_BT}\right)^{1/2}~\&~k_BT_0=\frac{18\alpha^3 }{N(E_F)}
\end{equation}
where $\nu_{ph}$ is the phonon frequency corresponding to the Debye temperature ($\sim10^{13}$~Hz), and $\alpha$ is the inverse localization length \cite{MottCP69}. For further analysis, we have plotted ln($\sigma T^{1/2}$) versus T$^{-1/4}$ in the 50--140~K temperature range and the obtained linear behavior manifests that the conduction in this range is governed by the Mott-VRH mechanism. By fitting a straight line to the experimental data we determine the VRH parameters $\alpha$, T$_0$, N(E$_F$), which are 1.3$\times$10$^{8}$~cm$^{-1}$, 1.6$\times$10$^{8}$~K, and 2.9$\times$10$^{21}$~eV$^{-1}$cm$^{-3}$, respectively. Furthermore, the values of the mean hopping distance (R) and mean hopping energy (W) can be calculated using $R=[9/8\alpha \pi kTN(E_F)]^{1/4}$~cm and $W= [3/4\pi R^3 N(E_F)]$~eV \cite{PaulPRL73}. 
The obtained values of R and W are 1.03~nm, and 76.2~meV, respectively. These values further support the model of Mott-VRH conduction, i.e., $\alpha R\textgreater\textgreater$1 and W$\textgreater\textgreater k_B$T \cite{PaulPRL73}, which infers that the conduction takes place via VRH in the low-temperature range. 

\begin{figure}[h]
\centering
\includegraphics[width=3.5in]{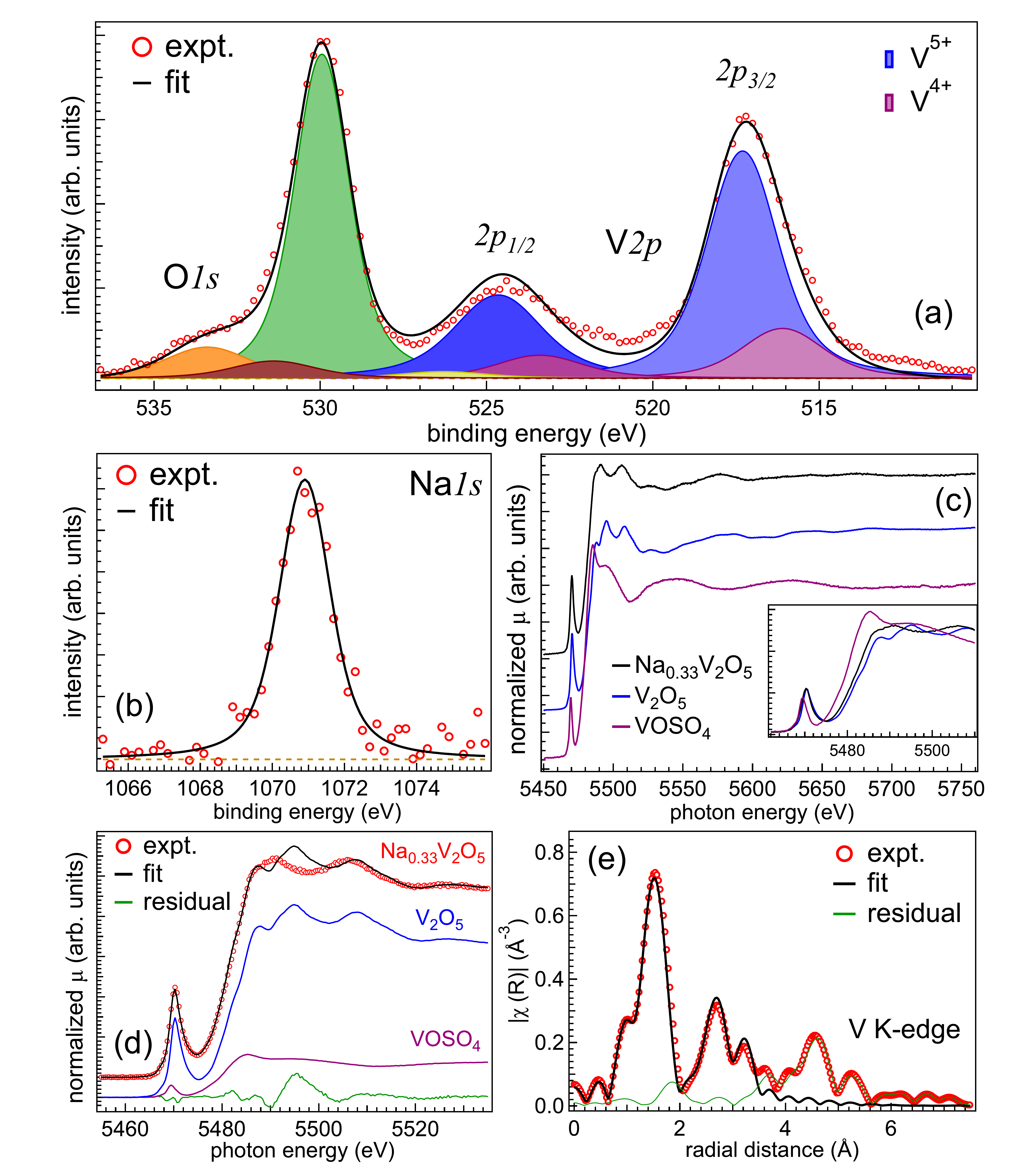}
\caption{(a) The O 1$s$ and V 2$p$, and (b) Na 1$s$ core-levels, (c) the XANES curves compared (vertically shifted for the clarity in presentation) with the references V$_2$O$_5$ and VOSO$_4$ samples for the 5+ and 4+ oxidation states, respectively; this comparison is further highlighted in the inset, (d) the linear combination fitting of the XANES spectra using the standard samples for 5+ and 4+ oxidation states, and (e) the Fourier transformed EXAFS curve of the $\beta$-Na$_{0.33}$V$_2$O$_5$ sample at the V K-edge where the open circles present the experimental data and the solid black lines are the best fits to the data.}
\label{fig:XPS+XAS}
\end{figure}

Finally, we use x-ray photoemission and x-ray absorption spectroscopy to understand the electronic properties and local structure of $\beta$-Na$_{0.33}$V$_2$O$_5$ at room temperature. Fig.~\ref{fig:XPS+XAS}(a) shows the O 1$s$ and V 2$p$ core-levels in the vicinity, where the spectral contribution due to Al-K$\alpha_3,\alpha_4$ satellites was subtracted following the reference \cite{Hufner03}. All the core-level spectra were calibrated with respect to the C 1$s$ peak at the 284.8~eV binding energy (BE). Further, we deconvolute these core levels using the Voigt function (a combination of Lorentzian and Gaussian lineshapes) after the subtraction of the Tougaard background. In the present case, the binding energy of O 1$s$ core-level is found to be $\sim$530~eV and an additional component at 532.6~eV is related to the chemically adsorbed oxygen in the form of C--O(H) \cite{SilverJESRP04}. The V 2$p$ core-level shows two spin-orbit split components having a separation of $\approx$7.35~eV and the absolute BE values are at 516.1 and 523.4~eV for V$^{4+}$, while 517.3 and 524.7~eV for V$^{5+}$ oxidation states, respectively, in good agreement with Refs.~\cite{Hufner03, SawatzkyPRB79, ZimmermannJPCM98, MendialduaJESRP95}. We also notice small shake-up satellite features around 14.1~eV and 10.2~eV from the 2$p_{3/2}$ components of V$^{5+}$, and V$^{4+}$ states, respectively. Moreover, a satellite at 8.3~eV is observed from the 2$p_{1/2}$ component of the V$^{4+}$ state. For the quantitative analysis, we compare the total area ratio of the V$^{5+}$ and V$^{4+}$ oxidation states and found that they exist in a percentage of 80\% and 20\%, respectively. This obtained ratio of the oxidation states is in agreement with the one found from the analysis of magnetization data. Fig.~\ref{fig:XPS+XAS}(b) shows} the Na 1$s$ core-level, where the observed BE value is 1071~eV, which confirms the Na in the 1+ oxidation state \cite{PrechtPCCP16}. 

Further, in Fig.~\ref{fig:XPS+XAS}(c) we present the near-edge x-ray absorption spectra (XANES) of the $\beta$-Na$_{0.33}$V$_2$O$_5$ sample and compare it with the corresponding spectra of the reference samples V$_2$O$_5$ (for V$^{5+}$) and VOSO$_4$ (for V$^{4+}$). The calibration was performed using the vanadium metal foil by fixing the derivative of first-order maxima at 5465~eV, and then shifting the spectra of $\beta$-Na$_{0.33}$V$_2$O$_5$ accordingly. Moreover, we subtracted the background from the recorded absorption spectra and then normalized it to unity above 600~eV from the edge jump \cite{ShuklaJPCC21}. The position of the absorption edge of the $\beta$-Na$_{0.33}$V$_2$O$_5$ sample (black solid line) is in between the two reference samples, i.e., V$_2$O$_5$ and VOSO$_4$ [see inset of Fig.~\ref{fig:XPS+XAS}(c)]. This indicates the presence of mixed 5+ and 4+ oxidation states of V ions in the $\beta$-Na$_{0.33}$V$_2$O$_5$ sample, consistent with the XPS and magnetization results (discussed above). A quantitative ratio of the mixed 5+ and 4+ states can be approximated from a linear combination fitting using the Athena software \cite{RavelJSR05}. Therefore, we have performed the linear combination fitting using the standard reference samples of VOSO$_4$ and V$_2$O$_5$ for the oxidation states of 4+ and 5+, respectively, and obtained a ratio of 83\% and 17\%, respectively [see Fig.~\ref{fig:XPS+XAS}(d)] for the $\beta$-Na$_{0.33}$V$_2$O$_5$ sample. This is again in good agreement with our XPS and magnetization results as well as with references \cite{BadotJSSC91, HeinrichPRL04}. The strong intensity pre-edge peak emerges due to the tetrahedral symmetry of present V-ions in the monoclinic phase  \cite{ChandraEA20, YamamatoXRS08}. Moreover, it is important to note here that if we consider the presence of three inequivalent absorbing sites of V atoms, in the anisotropic structure of $\beta$-Na$_{0.33}$V$_2$O$_5$ sample, the analysis of EXAFS curve becomes challenging due to increased number of scattering paths \cite{RavelJSR14}. On the other hand, in this line Ravel introduced a new formalism for such EXAFS analysis and applied for CaZrTi$_2$O$_7$ sample, which consists of three inequivalent Ti sites \cite{RavelJSR14}. In this paper, the complexity of the EXAFS curve fitting is minimized through the concept of fuzzy degeneracy in which the contributions of similar scattering paths are averaged in a given margin, known as the bin size, i.e., the scattering paths are grouped together (become degenerate) if they fall within this bin size \cite{RavelJSR14}. Interestingly, in this way, the use of fuzzy degeneracy reduces the number of scattering paths without affecting the validity/accuracy of the analysis \cite{RavelJSR14}. Moreover, in this formalism the obtained scattering paths become independent of the crystallographic orientations as the similar scattering paths are averaged over the inequivalent sites. This procedure is also applicable and relevant in the present case of inequivalent V sites in the $\beta$-Na$_{0.33}$V$_2$O$_5$ sample, as discussed above in XRD analysis (see Table II of \cite{SI}). 

\begin{table}[h]
 \caption{The EXAFS fitting parameters from the V K-edge of the $\beta$-Na$_{0.33}$V$_2$O$_5$ sample; path degeneracy, effective bond lengths (R$\rm_{eff}$), scattering paths, and the D-W factor.}
 \vskip 0.2cm
 \centering 
 \begin{tabular}{|c|c|c|c|}
    \hline 
 path degeneracy&scattering paths&~~R$\rm_{eff}~~$~&D-W factor\\
&&~(\AA)&(~\AA$^2$)\\
 \hline
 1.00&V--O (apical) &1.588&0.007\\
 \hline
 0.67&V--O (basal) &1.857&0.003\\
 \hline
 2.00&V--O (basal)&1.948&0.007\\
 \hline
 0.33&V--O (basal)&1.833&0.004\\
 \hline
 2.00&V--O (basal)&1.879&0.001\\
 \hline
 0.67&V--O (farthest)&2.315&0.006\\
 \hline
 1.33&V--V&3.005&0.006\\
 \hline
 2.33&V--V&3.602&0.015\\
 \hline
 \end{tabular}
 \label{Tab:EXAFS}
\end{table}

Therefore, for the analysis of the EXAFS at V K-edge, we follow the same formalism of fuzzy degeneracy (bin size of 0.03~\AA~ in the present case), and the path degeneracies are calculated over the different sites by the fractional population of the individual site in the unit cell \cite{RavelJSR14}. For example, the degeneracy having a value of 0.33 in Table III means that it includes only one scattering path. The structural parameters obtained from the Rietveld refinement of the XRD pattern, shown in Fig.~\ref{fig:XRD}(a), were used for the FEFF code of the Artemis program after the reduction and normalization of the absorption spectra in the Athena software available in the Demeter package \cite{RavelJSR05}. Then, we fit the Fourier transform of the EXAFS curve [k$^2$$\chi(k)$] in terms of $\chi(R)$ versus radial distance (R) up to 4.1~\AA~ with the single scattering paths \cite{AjayPRB22} from the absorbing V atoms (V--O and V--V) [see Fig.~\ref{fig:XPS+XAS}(e)]. The path degeneracy, effective bond lengths (R$_{\rm eff}$), and Debye-Waller (D-W) factor extracted for the scattering paths V--O and V--V are summarized in Table III. The obtained parameters indicate that there are three groups of V--O bond lengths, which are in good agreement with the bond length parameters extracted from the Rietveld refinement of the XRD pattern (see Table II of \cite{SI}). Interestingly, these three groups of bond lengths (apical, basal, farthest) were utilized for the in-depth understanding of Raman spectra, which establish the correlation between vibrational and electronic properties of the $\beta$-Na$_{0.33}$V$_2$O$_5$ sample.  

\section{\noindent ~Conclusions}

In summary, we have investigated the structural, magnetic, electronic, and lattice vibrational properties of $\beta$-Na$_{0.33}$V$_2$O$_5$ by performing a detailed analysis of various experimental results. The Rietveld refined XRD pattern confirms the monoclinic crystal structure with the space group C2/m (\#12). The Raman modes in a large temperature range of 13--673 K are assigned to different translational/chain rotations of Na and stretching/bending vibrations of different V-O bonds. Intriguingly, an anharmonic dependence of peak-positions and FWHM of the profound phonon modes is accredited to the symmetric phonon decay. Below about 40~K, a deviation in the phonon frequency and linewidth beyond the anharmonic fit is attributed to a spin-phonon interaction. The experimental value of $\mu_{\rm eff}$ manifests the presence of V-ions in mixed-valence states of 4+ and 5+ in a ratio of 18\% and 82\%, respectively, which is further confirmed by the XPS measurements. Also, a linear combination fit of the XANES at the V K-edge validates these observations. Interestingly, a detailed EXAFS curve fitting at the V K-edge provides insight about the local coordination and effective V-O bond-lengths, which are consistent with the one obtained from XRD analysis. Furthermore, the resistivity data demonstrates a semiconducting nature and follows three different conduction mechanisms in which Arrhenius type dominants in 370--290~K range (E$_a$=145~meV). The carrier conduction in the lower temperature range is dominated by the hopping mechanism, whereas the NNH model is followed in 260--190~K (E$_a$=92~meV), and then the 3D-VRH type of electrical conduction in 140--50~K region yields N(E$_{\rm F}$) = 2.9$\times$10$^{21}$~eV$^{-1}$cm$^{-3}$ and affirms the $\beta$-Na$_{0.33}$V$_2$O$_5$ sample as a Mott-insulator. 

\section{\noindent ~Acknowledgments}

RS thanks DST-Inspire, India for the fellowship, and the department of physics, IIT Delhi for providing XRD, PPMS and SQUID facilities. We thank Ravi Kumar and S. N. Jha for help and support during XAS measurements at RRCAT, India. RSD gratefully acknowledges the department of science \& technology (DST), India for support through Indo-Australia early and mid-career researchers (EMCR) fellowship (IA/INDOAUST/F-19/2017/1887) and UNSW for hosting his visit. We acknowledge the support from SERB-DST through core research grant (CRG/2020/003436). CU acknowledges the support of the Australlian Research Council through the Discovery Grant DP160100545. We also thank Yousef Kareri for help during Raman spectroscopy measurements at UNSW, Australia.

\end{document}